\documentclass{deepmind}

\title{Launchpad: A Programming Model for Distributed Machine Learning Research}

\author[1*]{Fan Yang}
\author[1*]{Gabriel Barth-Maron}
\author[2*]{Piotr Stańczyk}
\author[1*]{Matthew Hoffman}
\author[1]{Siqi Liu}
\author[1]{Manuel Kroiss}
\author[1]{Aedan Pope}
\author[1]{Alban Rrustemi}
\affil[1]{DeepMind}
\affil[2]{Google Brain}
\affil[*]{corresponding authors}

\usepackage[utf8]{inputenc}
\usepackage{placeins}
\usepackage{natbib}
\usepackage{graphicx}
\usepackage{todonotes}
\usepackage{listings}
\usepackage{tcolorbox}
\usepackage{enumitem}
\usepackage{subcaption}
\usepackage[newfloat,frozencache]{minted}
\usepackage{xspace}

\newmintinline[py]{python}{}
\newminted[python]{python}{fontsize=\small,frame=single}

\newcommand{\lstname}{Listing}
\SetupFloatingEnvironment{listing}{name=\lstname}

\newcommand{\Program}{\py{Program}\xspace}
\newcommand{\Node}{\py{Node}\xspace}
\newcommand{\Handle}{\py{Handle}\xspace}

\begin{abstract}
A major driver behind the success of modern machine learning algorithms has been their ability to process ever-larger amounts of data. As a result, the use of distributed systems in both research and production has become increasingly prevalent as a means to scale to this growing data. At the same time, however, distributing the learning process can drastically complicate the implementation of even simple algorithms. This is especially problematic as many machine learning practitioners are not well-versed in the design of distributed systems, let alone those that have complicated communication topologies.
In this work we introduce Launchpad, a programming model that simplifies the process of defining and launching distributed systems that is specifically tailored towards a machine learning audience. We describe our framework, its design philosophy and implementation, and give a number of examples of common learning algorithms whose designs are greatly simplified by this approach.
\end{abstract}

\begin{document}
\maketitle
\suppressfloats

\section{Introduction}

Modern numerical frameworks---e.g.\ TensorFlow, PyTorch, and JAX\ \citep{abadi-2015-tensorflow, jax2018github, paszke-2017-pytorch}---have significantly contributed to recent advances in machine learning. Perhaps \emph{the} key factor underlying the success of these tools is their ability to define a graph of the operations involved for a given numerical computation. This capability has proven powerful precisely because it enables the graph to be automatically differentiated, and as a result the frameworks are tailor-made for algorithms which revolve around consuming data and updating parameters using some form of gradient descent. An equally important attribute of these frameworks, however, lies in their use of operations which allow researchers to parallelize computation in a seamless manner. Such approaches allow researchers to greatly scale up model training both in terms of the amount of data able to be processed as well as the size of models themselves. In combination these two capabilities have directly led to numerous advances in the field \citep[e.g.][]{krizhevsky-2012-alexnet, brock-2018-biggan, silver-2018-alphazero}. 

At the same time, as both model and dataset sizes have grown, the distribution of computation across multiple devices has become increasingly common. Writing code that spans more than one device, though, introduces a new dimension of complexity---distributed systems and communication---that many machine learning practitioners are unfamiliar with. While the above mentioned frameworks do provide mechanisms for distribution, the focus on automatic differentiation can make such approaches cumbersome where differentiation through the communication channel is not necessary. Further, while such approaches are well-suited to settings where batches of incoming data can be partitioned and assigned to different devices, frequently this form of distribution can leave the communication between devices \emph{implicit} which can obscure program flow where more complex communication is required. This is especially apparent in settings such as reinforcement learning where the data generation process itself can frequently involve heterogeneous components with more complex computational requirements.

As an alternative to the implicit communication mechanisms of modern numerical frameworks one can instead rely on low-level, distributed communication protocols such as message passing \citep[e.g.\ MPI,][]{gropp-1999-mpi} or a remote procedure call (RPC) mechanism \citep{birrell-1984-rpc}. Modern implementations of such protocols include gRPC~\citep{grpc}, ZeroMQ~\citep{zeromq}, etc. A distributed learning algorithm can then be implemented using any of the aforementioned computational frameworks and connected using a suitable communication strategy. This can often prove less constraining than using a large computational framework for communication, since arbitrary messages can be passed between remote services which are no longer bound by implementing communication within the framework's domain specific language. One downside of this approach, however, is that it can often leave the definition of the communication topology to be defined in an ad-hoc basis by each individual service of the distributed algorithm.

\begin{figure*}
    \centering
    \begin{subfigure}{0.47\textwidth}
    \begin{python*}{gobble=8}
        # Create an empty program graph.
        p = Program('producer-consumer')
        
        # Add nodes producing a range of data.
        with p.group('producer'):
          h1 = p.add_node(RangeNode(0, 10))
          h2 = p.add_node(RangeNode(10, 20))
        
        # Add a node to consume from producers.
        with p.group('consumer'):
          p.add_node(ConsumerNode([h1, h2]))
    \end{python*}
    \end{subfigure}
    \hspace{.03\textwidth}
    \begin{subfigure}{0.47\textwidth}
        \includegraphics[width=\textwidth]{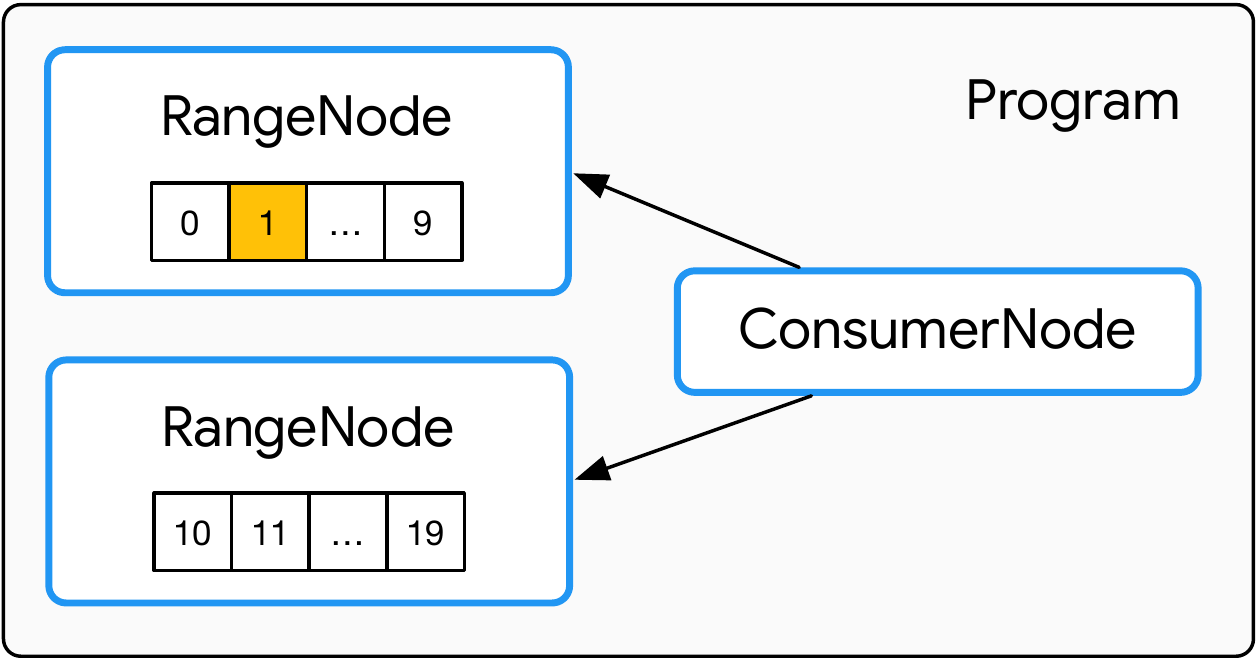}
    \end{subfigure}
    \caption{(left) A simple example Launchpad program and (right) the graph associated with this program. Although the implementation of these nodes is not shown, conceptually they correspond to two nodes which produce sequential data on request from a given range and a consumer node which performs some calculation on these inputs.}
    \label{fig:simple_program}
\end{figure*}

To address this problem, in this work we introduce Launchpad, a programming model that simplifies the process of defining and launching instances of distributed computation. The fundamental concept of this model is that of a Launchpad \emph{program} which represents computation as a directed graph of service \emph{nodes}. Edges in this graph denote communication between nodes which are made via remote procedure calls. By making the graph representation explicit Launchpad makes it easy to both design and later modify a program's topology in a single, centralized way---something that would not be possible were the system defined in a more ad-hoc fashion. Launchpad also clearly separates the program definition, as given by the graph datastructure, from the mechanism used to launch the distributed program. In particular this design choice allows for the same distributed system to be executed on different platforms (e.g.\ a single machine, cloud provider, or self-hosted cluster) by using different launcher implementations.

At launch time the program datastructure is translated into a collection of services, implemented as threads or processes on one or more machines. The launching mechanism is also responsible for setting up the communication channels for each service, as represented by edges in the graph. In particular we make use of a layer on top of gRPC~\citep{grpc} that we will detail in Section~\ref{sec:services}. After starting the services and initializing the communication channels the launcher mechanism runs the active computational process (defined by the service) or puts the service into a wait-loop to allow it to respond to requests. As a result, after launching a distributed program Launchpad adds no additional overhead as communication between individual services will be just as fast as the underlying communication protocol.

In the rest of this work we will detail the programming model, along with further details of its implementation and the life-cycle of a Launchpad program, in Sections~\ref{sec:programming_model}--\ref{sec:services}. To show the value of this approach, in Section~\ref{sec:examples} we will also give a comprehensive set of example machine learning algorithms designed using Launchpad, and show that this greatly simplifies their implementation. In this final section we will focus on the expressiveness of this framework. While we will give some information as to the computational efficiency of this approach we note that this is largely controlled by the efficiency of the underlying communication mechanism and the computation performed by the nodes themselves.

\subsection{Related work}

Ray~\citep{moritz-2017-ray} similarly provides a programming interface which decouples the task and the process boundaries. Ray is able to schedule a large number of Python tasks onto distributed processes and in addition, it provides fault tolerance and exact recovery mechanisms. However the use of implicit process boundaries can introduce additional difficulties when precise control of placement is required (e.g., performance sensitive tight loop), and the overhead of exact recovery is often unnecessary for machine learning applications. Additionally these recovery mechanisms often come at the price of extra complexity in the underlying systems, adding to the complexity necessary to debug and extend the system.

Another common approach to this problem involves drastically restricting the domain of target solutions. Examples of this include RLgraph and RLlib in the reinforcement learning setting \citep{liang-2017-rllib, schaarschmidt-2018-rlgraph}; Caffe and Keras for supervised learning \citep{jia-2014-caffe, chollet-2015-keras}. Unlike Launchpad and Ray these do not provide a generic interface for distributed computation, however they are often built on such a generic framework. For example RLlib builds on Ray, RLgraph and Keras are built on top of distributed TensorFlow (although these examples can also use a PyTorch backend).

\section{The Launchpad Programming Model}
\label{sec:programming_model}

Launchpad is a programming model that represents a distributed system as a graph datastructure describing the system's topology. Each \emph{node} in the graph represents a service in the program, i.e.\ nodes represent the fundamental units of computation that we are interested in running. Nodes themselves are datastructures which define the computation that \emph{will be run} for each service; in Section~\ref{sec:services} we will describe node types in more detail, however for the time-being they can be thought of as \emph{factory methods} for the actual computation. As nodes are added to the Launchpad graph, a \emph{handle} is constructed which acts as a reference to that node and ultimately represents a client to the yet to be constructed service. A directed edge in the program graph, representing communication between two services, is created when the handle associated with one node is given to another at construction time. This edge originates from the receiving node, indicating that the receiving node will be the one initiating communication. This allows Launchpad to easily define cross-service communication simply by passing around node handles.

Figure~\ref{fig:simple_program} provides an example of a simple Launchpad program and its associated graph. Here we can also see that node instances are typed, where the exact form of computation that will be executed depends on the \emph{service type} that is used. While the nodes constructed in this example are trivial and not indicative of actual node types, we will show that the mechanism for constructing and connecting these nodes enables many common usage patterns. In Section~\ref{sec:services} we will give more details on actual service types and detail a generic type that is useful more broadly; Section~\ref{sec:examples}, meanwhile, will give examples using these builtin types.
Nodes can also be grouped into disjoint sets to form what we refer to as \emph{resource groups}. Resource groups allow the program to set platform specific resource constraints (e.g.\ the amount of RAM to allocate or number of accelerators to use) at launch time\footnote{The open-sourced version of Launchpad omits the setting of resource constraints as currently released launchers are restricted to single-machine settings.}. These resource specifications will be applied to each node in a group's set. By definition, each node can only be assigned to one resource group, and all unassigned nodes will be placed into a default resource group.

While the program datastructure is ultimately responsible for describing the distributed system and its layout, when launching and executing the system the program delegates to each individual node (itself an instance of a node or \emph{service type}) in order to run the service's computation. We will refer to these individual units of computation as \emph{executables} in order to emphasize the fact that a service itself may be made up of multiple executables. By decoupling the declaration of a node from its associated implementation we are able to improve the flexibility of the systems generated by Launchpad: the materialized executables of a node could be a process, a set of processes, or even just referencing existing services. In Section~\ref{sec:services} we give a more thorough description of service types along details pertaining to a few common types. Finally, Launchpad provides platform-specific \emph{launchers} which are responsible for processing the program in order to create and compile all of the worker objects, allocate resources and constraints, and ultimately for sending the executables to the platform and executing them. By separating the implementation of the program logic from the launching and execution logic we are able to both create distributed programs that are capable of running on multiple platforms as well as to build launchers for novel platforms.

In what follows we will give a more detailed description of the internal implementation of Launchpad and in particular we will describe the full life-cycle of a program object from its creation to its launching and executation on a target platform.

\section{Program Life Cycle}
\label{sec:life_cycle}

The life cycle of a Launchpad program can broadly be broken into three phases: \emph{setup}, \emph{launch}, and \emph{execution}. The setup phase involves construction of the program datastructure. In the launch phase this datastructure is processed in order to allocate resources, addresses, etc. as well as to initiate launch of the specified services by performing any necessary setup for the executables that will be run. Finally, the execution phase begins running the services, including creating any necessary clients for communication between services. From the perspective of Launchpad the execution phase is the final phase, after which all control has been handed over to the individual executables.

In what follows we will describe each of the three phases in more detail and outline the role they play within the life cycle of a program. 
We also note that Launchpad has been implemented in Python due to that language's prevalence in scientific computation, and this provides a simple language in which to define the program and node datastructures and launchers for individual platforms. The design of this framework, however, could easily be implemented in any other host language, and as we will later see in Section~\ref{sec:services} this does not prohibit services whose executables are implemented in a more low-level language, e.g.\ C/C++.

\subsection{Setup}
\label{sec:life_cycle:setup}

During setup the user first constructs a \Program object which corresponds to an empty graph. The \py{add_node} member function can then be used to incrementally add nodes to this graph. This method takes a \Node instance, adds it to the internal graph structure, and returns a \Handle which acts as a reference to the node. Internally, each node exposes a \py{create_handle} method which allows the node instance to create a service-specific handle. As described earlier, this handle can be passed to other nodes in order to designate a communication channel. During the execution phase, each handle will be dereferenced to return a service-specific client implementing this communication. The exact form of client returned will depend on the service type, which we will describe in more detail in Section~\ref{sec:services}.

The setup phase is also when the user can create resource groups and assign individual nodes to these groups. This is implemented as a context created by the \py{Program.group} method, such that any node added under this context is assigned to the group. This grouping is useful for applying a common resource definition to a set of nodes sharing homogeneous resource requirements, for example, a set of data-generating nodes in reinforcement learning may share the same requirements for each simulator.
For simplicity we also require that nodes added to the same resource group share a node type; this conceptually simplifies the process by requiring nodes to have suitably comparable executables. In this phase it is worth noting that groups are only created---actual constraints and requirements are not assigned until the next phase (the launch phase) so that requirements can be platform-specific.

The earlier listing in Figure~\ref{fig:simple_program} provides an example of this setup phase for a simple program. In this example two nodes are created and added under the \py{'producer'} resource group and connected to a single node in the \py{'consumer'} group. As we can see from this example, the setup phase can best be summarized as the ``user facing'' portion of a program's life cycle, wherein the graph structure itself is generated before passing it to the launch phase which we will describe next.

Finally, although the setup phase is mostly concerned with describing the structure of a program, there is also some limited internal work necessary to enable the later phases. In particular, in order to allow communication between different services during execution, the \Handle object associated with a given node needs an address (e.g.\ an IP address) pointing to the service represented by each conceptual \Node. However, executables of a node are not created nor are the associated addresses allocated until the launch phase---it is important to note that physical addresses are platform specific and hence unavailable during setup. Instead, the \Node instance must create address placeholders and assign them to its associated \Handle. These placeholders will then be filled in during the launch phase.

\subsection{Launch}
\label{sec:life_cycle:launch}

The launch phase begins when a \Program instance is passed to a platform-specific launcher. An optional mapping from resource group identifiers to platform-specific resource requirements may also be passed to the launcher. These requirements will be used by the launcher when allocating resources for the underlying executables it creates. As noted earlier, the currently open-sourced version of Launchpad does not make use of these resource requirements as the released launcher launches and executes the program within a single machine.
In \lstname~\ref{lst:launcher} we show an example launcher used to run a program with both resource groups and colocations. We also set at launch time a collection of resource requirements (which will be interpreted by the launcher).

\begin{listing}
\begin{python}
# Define resources.
resources = {
  'producer': {'cpu': 2, 'ram': 2},
  'consumer': {'cpu': 2, 'ram': 2, 'gpu': 1}
}

# Launch the program with assigned resources.
launch(p, resources)
\end{python}
\caption{Example usage of resource constraints used at launch time.}
\label{lst:launcher}
\end{listing}

Upon receiving the program, the launcher must first contact the corresponding distributed platforms for resource discovery and provisioning. The launcher can then examine the handles referenced in the program and assign platform-specific physical addresses to the associated address placeholders in a manner that enforces the constraints imposed by the relevant resource groups.
As handles may be arbitrarily distributed among the collection of workers, rather than individually converting address placeholders to physical addresses the launcher constructs an \emph{address table} to hold this mapping. This will be used during the execution phase when the handles are dereferenced.

Next, the launcher iterates over all nodes in the \Program and calls the \py{to_executables} method on each node to create the backing executables. Since all the addresses are resolved at this point, if a node receives handles, its corresponding executables will be able to contact the services using the resolved addresses correctly. The implementation \py{to_executables} has access to both the launch type and any associated resource specifications, which it can use to generate the most suitable executables.

After necessary compilations, all the generated executables are then sent to the corresponding distributed platform for execution.
Optionally, the launcher can wait for or monitor the individual nodes after they begin execution. This is especially useful in integration tests (distributed or single process) in which we want to verify that the distributed system (specified by the \Program) performs a task and terminates correctly.

\subsection{Execution}
\label{sec:life_cycle:execution}

While the program is defined and launched in a centralized way, the execution is completely decentralized---and asynchronous in general. The individual hosts in the target platform will start executing the executables once they receive them. Services will be set up and bound to the addresses determined at the launch phase, so that they can contact and interact with each other.

In particular, execution begins by running all of the executables created during the launch phase. Each executable enters its \py{run} method which is responsible for setting up any communication clients associated with the addresses (defined via \Handle{}s) generated during the previous phase. The executable can then proceed to set up and run its service, at which point all computation is in the hands of the given service type.

\section{Service Types}
\label{sec:services}

The computational building blocks exposed by Launchpad are represented by different service types, where again each service type is represented by \Node and \Handle classes specific to that type. The \Node class is the user-facing interface to a given service type for configuring specific services. While \Handle instances are also exposed to the user---and passing them between nodes is used to designate communication channels---their primary use is via the \py{dereference} method used to create the actual client objects. Having a \Handle abstraction over the actual client object not only allows us to define a complete program before platform-specific addresses are available in the setup phase, but also allows us to flexibly choose the most appropriate client type at launch phase (e.g., to use a shared-memory channel if the service is allocated on the same physical machine).

Nodes communicate with other nodes by means of the clients returned by dereferencing each \Handle object. While we make no restrictions on the form of the client object created by dereferencing, in many cases clients are exposed as arbitrary Python objects whose member functions correspond to remote procedure calls (RPC) to another service. Individual node instances may make more restrictive assumptions as to the form of clients they accept, however this is left to the discretion of that particular service type.

\subsection{Python Nodes}
\label{sec:services:py_node}

It is possible to implement individual service types for each computational unit, however this would quickly become burdensome. Instead, the most common use case is to use Python objects to represent each service. From a computational perspective the use of Python is not an impediment to speed or scalability as in most, if not all, machine learning applications any computationally intensive services will make use of a lower-level computational framework such as TensorFlow, PyTorch or JAX.

Launchpad exposes services of this form by introducing \py{PyNode} and \py{CourierNode} classes. Both classes take as input a Python class and arguments to the class's constructor such that instances of these nodes act as \emph{deferred constructors} for the underlying class. Additionally any input arguments that are given as \Handle{}s to other nodes will treated specially so that communication clients can be created when the nodes are executed. The difference between these two node types is that the \py{PyNode} type \emph{does not} return a \Handle and as a result cannot receive messages---this is implemented purely as a cost saving procedure that can be used when the program designer knows that this node instance is purely used for execution or for initiating communication. As a result, we will focus on the \py{CourierNode} type\footnote{These nodes are named for Launchpad's internal communication library, Courier, which is built on top of gRPC} which does expose a \Handle reference.

Handles created by \py{CourierNode} instances are ultimately exposed as generic RPC objects, which means that from the perspective of any consuming class remote communication is invisible and it appears as if it is just using the original Python objects. Usage of this service type is shown in \lstname~\ref{lst:producer-consumer}, which shows a full implementation of the example given in Figure~\ref{fig:simple_program}. Note that the constructed Python object is not given to \py{CourierNode}. This is because we make use of the ability to serialize the class and its arguments and constructing the object may have undesirable side-effects that we wish to avoid during the launch phase.

Along with this node are corresponding handle and executable implementations. Construction of the Python class given to the node instance is deferred at launch time, and in this case the \py{CourierNode} implementation serializes the class and any given argument, which are then shipped over network and deserialized at execution time. Since the arguments to the node can include handle instances, it is during this deserialization that the worker will also dereference any handle instances included before using them to construct the given class. The worker then starts an RPC server which exposes all public methods from the class save for a \py{run} method if it exists. If such a \py{run} method exists the worker will execute it, otherwise it will wait on any incoming RPC invocations. The associated \py{CourierHandle} similarly implements, in its \py{dereference} method the construction of an RPC client which exposes each of the public methods of the associated class. 

As noted earlier the RPC mechanism we build upon uses the low-level GRPC library. However, at a high level our main requirements are the ability to expose an arbitrary Python object as a server and to construct a client that enables calls to this remote object. To that end there are a number of publicly available frameworks\footnote{Examples include xmlrpc (standard library), zerorpc (\url{www.zerorpc.io}), or Pyro4 (\url{pythonhosted.org/Pyro4}).} that could also be used.

Due to its generality this node type forms the basis of most of the examples we will give in Section~\ref{sec:examples} as well as many of the use cases we have seen in practice. However, there do exist uses that may require more specialized or low-level implementations that we will briefly describe in the next section.

\begin{listing}
\begin{subfigure}[t]{0.45\textwidth}
\begin{python}
class Range:
  def __init__(self, s, e):
    self._it = iter(range(s, e))
    self._size = e - s

  def get_size(self):
    return self._size
  
  def produce(self):
    return next(self._it)

class Consumer:
  def __init__(self, producers):
    self._producers = producers
    
  def run(self):
    for p in self._producers:
      for _ in range(p.get_size()):
        print(p.produce())
\end{python}
\end{subfigure}
\hspace{0.01\textwidth}
\begin{subfigure}[t]{0.53\textwidth}
\begin{python}
# Create an empty program graph.
p = Program('producer-consumer')

# Add nodes producing a range of data.
with p.group('producer'):
  r1 = p.add_node(CourierNode(Range, 0, 10))
  r2 = p.add_node(CourierNode(Range, 10, 20))

# Add a node to consume from producers.
with p.group('consumer'):
  p.add_node(CourierNode(Consumer, [r1, r2]))
\end{python}
\end{subfigure}
\caption{A simple producer-consumer example with two services producing values and one service that consuming those values; this is a full implementation of the example from Figure~\ref{fig:simple_program} using \mintinline{python}{CourierNode} instances.}
\label{lst:producer-consumer}
\end{listing}

\subsection{Other Node Types}
\label{sec:caching-service}

\paragraph{Caching} Caching is commonly used in order to reduce the computational burden introduced by frequent communication. For example, in Section~\ref{sec:parameter_server} we introduce a parameter server model (itself a form of caching) and show how to further reduce the burden placed on the server by many incoming requests. In \lstname~\ref{lst:caching} we use a simple Python \py{Cacher} implementation which does exactly this. However, as caching reoccurs quite frequently we also expose a generic, low-level \py{CacherNode}. This node takes as input any other \py{CourierNode} instance and caches the results of RPC calls to this node for a given amount of time. Caching exists as a good example of a service type within Launchpad that is generically useful and for which it is beneficial to have a high-performance low-level implementation.

\paragraph{Colocation} In a heterogeneous computational environment another constraint of great importance is the relative locality of different services. For example, it may be desirable that two services are colocated on the same physical machine as threads or processes. To enable this type of control, we introduce a special \py{ColocationNode} type that wraps a set of other nodes. At execution time the executable generated by this node will instantiate the wrapped nodes using either local processes or threads. This enables the program designer to more carefully control the locality and speed of communication on a node-by-node basis.

\paragraph{Data services} A numerical process that samples from a dataset in order to optimize a loss is an incredibly common computational template in machine learning. A specialized node exposing such a dataset may not be necessary as long as the dataset can fit into memory and/or be read from disk. However in situations where this is not the case Launchpad also provides a \py{ReverbNode} which can access a dataset provided by Reverb~\citep{cassirer2021reverb}. This is particularly useful in reinforcement learning settings where the dataset can itself be filled in an online fashion by data generating processes that interact with and explore an environment. We will see a further example of this in Section~\ref{sec:examples:rl}.

\section{Examples}
\label{sec:examples}

Launchpad's programming model is rich enough to represent a wide variety of distributed systems. We demonstrate this by introducing several examples that exemplify some common distributed systems paradigms. Note that all of these examples use the generic \py{CourierNode} from Section \ref{sec:services:py_node}, but it is also possible to implement specialized node types instead.

\subsection{Parameter Server}
\label{sec:parameter_server}

Parameter servers \citep{ahmed-2012-paramserver, li-2014-paramserver} are an approach for addressing the storage and updating of parameters in distributed machine learning systems. Typically this is used in systems where there are one or more processes updating model parameters in parallel, each of which periodically synchronizes their parameters with the centralized server.
In \lstname~\ref{lst:parameter-server} we consider a simplified variant of this problem wherein there is a single parameter server and multiple external services requesting updated parameters from this server. This example includes both the underlying python implementations of the servers, as well as the Launchpad program which connects these components together. For the sake of brevity the server in this example returns random values, but it would be straightforward to extend this to meaningful parameters. Here we are primarily interested in the form that communication might take in this setting. Note also that this is a common occurrence in distributed reinforcement learning algorithms as we will see in Section~\ref{sec:examples:rl}.  

\begin{listing}
\begin{subfigure}[t]{0.45\textwidth}
\begin{python}
class ParamServer:
  def get_value(self):
    # Sleep for 1ms to simulate delay
    # in retrieving parameters.
    time.sleep(0.001)
    return random.random()

class Requester:
  def __init__(self, param_server):
    self._param_server = param_server

  def run(self):
    while True:
      v = self._param_server.get_value()
      print('Received: {}'.format(v))
\end{python}
\end{subfigure}
\hspace{.01\textwidth}
\begin{subfigure}[t]{0.53\textwidth}
\begin{python}
# Create an empty program graph.
p = Program('ps')

# Add a node for the parameter server.
with p.group('server'):
  server = p.add_node(CourierNode(ParamServer))

# Add nodes for the requesters.
with p.group('requester'):
  for _ in range(num_requesters):
    p.add_node(CourierNode(Requester, server))
\end{python}
\end{subfigure}
\caption{A single parameter server that serves parameters to multiple requesters.}
\label{lst:parameter-server}
\end{listing}

Interestingly, this example allows us to discuss issues that occur when scaling distributed systems with fan-in topologies. In particular we can see how Launchpad allows us to easily implement solutions by modifying the system's topology. Consider, for example what happens to this program as we increase the number of requesters. We can see that the server will not be able to handle requests fast enough, which slows the entire system down and prevents it from scaling effectively. One way of making the system more scalable is to replicate the servers and partition the requesters among them. In Launchpad this is as simple as adding more \py{Server} nodes to the program and distributing the requesters evenly among them. We can see an example of this in the (left) of \lstname~\ref{lst:caching}.

\begin{listing}
\begin{subfigure}[t]{0.49\textwidth}
\begin{python}
# Create an empty program graph.
p = Program('ps')

# Add nodes for parameter servers.
with p.group('server'):
  servers = [
    p.add_node(CourierNode(Server))
    for _ in range(num_servers)
  ]

# Servers have 1/num_servers requesters.
with p.group('requester'):
  for i in range(num_requesters):
    server = servers[i % num_servers]
    p.add_node(
      CourierNode(Requester, server))
\end{python}
\end{subfigure}
\hspace{.01\textwidth}
\begin{subfigure}[t]{0.49\textwidth}
\begin{python}
# Create an empty program graph.
p = Program('ps')

# Add a node for the parameter server.
with p.group('server'):
  server = p.add_node(Server)

# Add a node for the caching layer.
with p.group('cacher'):
  cacher = p.add_node(
    CourierNode(Cacher, server, timeout))

# Requesters use the caching layer.
with p.group('requester'):
  for _ in range(num_requesters):
    p.add_node(
      CourierNode(Requester, cacher))
\end{python}
\end{subfigure}
\caption{(left) The parameter server can be replicated across many instances so it can scale to handle many requesters. Each instance is used to serve parameters to only a fraction of the requesters. (right) Another way to help the parameter server scale to many requesters is to introduce a caching layer can be introduced between it and the requesters. The requesters now receive (potentially stale) values from the cacher.}
\label{lst:caching}
\end{listing}

Alternatively, the original implementation can be made more scalable by introducing a caching layer between the server and requesters; this is shown in the (right) of \lstname~\ref{lst:caching}. Here requesters retrieve values from a cacher instead of communicating directly with the server. If the cacher has a fresh copy of the value (freshness is typically determined by a timeout parameter) then it will return that value without making a request to the server. Otherwise the cacher will request a new copy of the value from the server, store it, and return it to the requester. See \lstname~\ref{lst:pycacher} in the Appendix for an implementation of the cacher (although as noted earlier Launchpad also provides a low-level implementation of this procedure).

\begin{figure}
\centering
\includegraphics[width=0.6\textwidth]{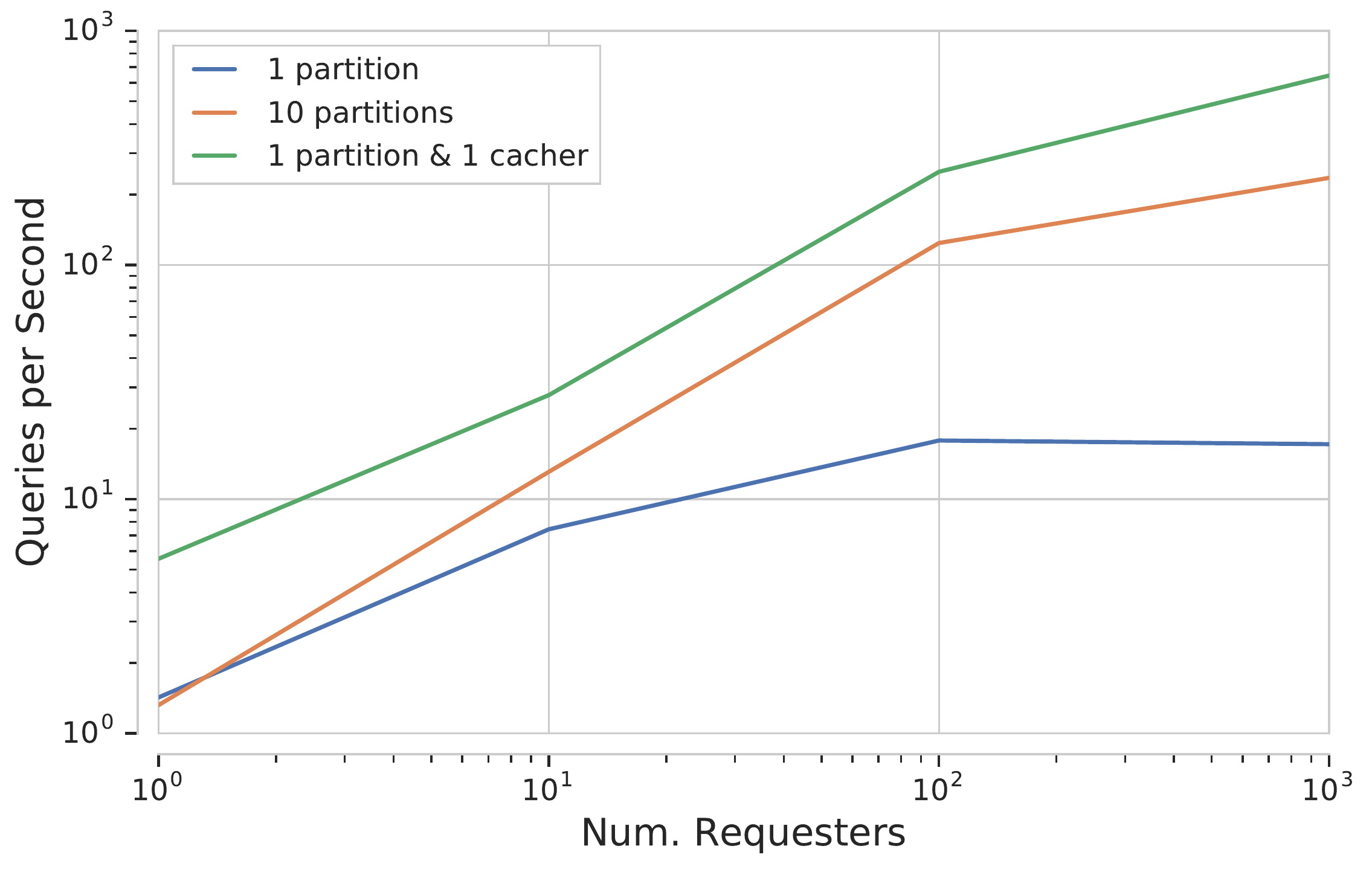}
\caption{Performance of various implementations of a distributed parameter server. Here we see the best performance and simplest implementation for the single partition/caching approach. While this may differ based on various implementation characteristics the approach to testing this procedure is drastically simplified by using Launchpad to test various approaches.}
\label{fig:ps-performance}
\end{figure}

In Figure~\ref{fig:ps-performance} we show example behavior of applying these different techniques to improve the performance of this system as we increase the number of requesters. Shown is the performance in terms of queries per second (QPS), where we have scaled the initial performance to be 1 QPS. In particular we see that the caching layer offers us the best performance tradeoff, however, more important than this insight, is the simple way in which we are able to prototype these improvements with Launchpad.
These two approaches to scaling the Parameter Server framework offer different trade-offs and their performance will change for different scenarios. Because Launchpad allows users to easily modify our implementation they can easily compare both and use whichever is best for their use case.

\subsection{MapReduce}
\label{sec:map-reduce}

MapReduce is a widely used model for processing large data sets \cite{dean-2008-mapreduce}. It consists of map functions that process data to output intermediate key/value tuples, and reduce functions that aggregate them. These map and reduce functions are typically parallelized by distributing them across many machines. In \lstname~\ref{lst:mapreduce} we use Launchpad to implement MapReduce. Although typically one would use a specialized framework for this type of computation, here we show a proof of concept which could be used as a prototype or combined with other Launchpad nodes.

The example we give counts the frequency of words found in a given set of files. Our map function \py{WordMapper} takes an input text file, splits its content into word tokens, and sends these words to a reducer; the reducer, implemented as \py{CountReducer}, keeps count of how often has seen each word. See \lstname~\ref{lst:mapper-reducer} in the Appendix for implementations of \py{WordMapper} and \py{CountReducer}. Our Launchpad program consists of one \py{WordMapper} node per input file, and as many \py{CountReducer} nodes as necessary.

\begin{listing}
\begin{python}
# Create an empty program graph.
p = Program('mapreduce')

# Define paths to input and output files.
out_path = '/some/dir/outfile.txt'
in_paths = ['/some/dir/infile1.txt', '/some/dir/infile2.txt']

# Add nodes for the reducers. These will
# write their output to out_path.
reducers = []
with p.group('reducer'):
  for _ in range(num_reducers):
    reducers.append(p.add_node(CourierNode(CountReducer, out_path)))

# Add nodes for the mappers.
with p.group('mapper'):
  for path in in_paths:
    p.add_node(CourierNode(WordMapper, path, reducers))
\end{python}
\caption{A MapReduce program for counting the frequency of words in text files. One map node is created for each input file, and these are connected to several reducer nodes that count the word frequencies.}
\label{lst:mapreduce}
\end{listing}

\subsection{Evolution Strategies}
\label{sec:evolution}

Evolution Strategies (ES) represent a collection of black-box optimization techniques that optimize a parametric search distribution in order to maximize a given fitness function \citep[e.g.][]{ros-2008-cmaes, wierstra-2008-nes}. Such algorithms evaluate the fitness function on $N$ samples from the search distribution, and update the distribution based on these observed fitnesses. In this example we implement ES by separating the algorithm into an \py{Evaluator} class that takes the sampled parameters and evaluates their fitness and an \py{Evolver} class that maintains the search distribution, passing a single sampled set of parameters to each evaluator. Upon retrieving the fitness values the \py{Evolver} performs its update on the search distribution. See \lstname~\ref{lst:evolver} in the Appendix for example implementations of the \py{Evaluator} and \py{Evolver}.

Evaluating the fitness of the sampled parameters may be a computationally expensive operation, which is made worse by the fact that we want to perform $N$ evaluations. We can parallelize this computation by using Launchpad to setup a distributed system that has each of our evaluators and the evolver run in their own service, in particular this allows us to run the evaluators in parallel on separate machines. We could also allow each evaluator access to a hardware accelerator (typically a GPU) to provide further speed improvements should the underlying fitness function be computationally taxing.

The Launchpad \Program we create in \lstname~\ref{lst:nes} has one node for each of the $N$ evaluators and another node for the evolver. The evolver node receives handles to all of the evaluators so it can send them sampled parameters and receive fitness measurements. Note that, for this implementation, we obtain futures from the evaluators when they calculate the fitness. Doing so allows the \py{Evolver} to make requests to all of the evaluators at the same time and then wait for them all to finish calculating the fitness values.

\begin{listing}
\begin{python}
# Create an empty program graph.
p = Program('es')

# We use a resource group to set evaluator
# specific resources at launch.
with p.group('evaluator'):
  # Add nodes for fitness evaluators.
  evaluators = [p.add_node(CourierNode(Evaluator)) for _ in range(num_evaluators)]

# Add a node for the evolver.
with p.group('evolver'):
  p.add_node(CourierNode(Evolver, evaluators))
\end{python}
\caption{A Launchpad program for Evolution Strategies. Multiple evaluators are added within a single resource group. This resource group can be referred to at launch time to set platform-specific resources for the evaluators (i.e.\ each evaluator may be given a GPU). All of the evaluators are passed to the evolver, which will use them for calculating fitness.}
\label{lst:nes}
\end{listing}

\subsection{Reinforcement Learning}
\label{sec:examples:rl}

A common program topology used in distributed reinforcement learning is the so-called \textit{actor-learner} architecture \citep{mnih-2016-a3c,espeholt-2018-impala}. This topology separates acting (generating actions with a policy and receiving observations from an environment) from learning (policy improvement), and uses multiple actors to parallelize data collection. The actors send trajectories of generated data to the learner, and periodically fetch the latest variables from it. The learner (typically using one or more hardware accelerators) computes and optimizes losses (e.g.\ via policy gradients) on the batch of trajectories and updates the parameters via stochastic gradient descent. See \lstname~\ref{lst:actor-learner-program} for pseudo-code for how this might be implemented in Launchpad and \lstname~\ref{lst:actor-learner} in the Appendix for an example implementation of the \py{Actor} and \py{Learner}.

Further, complete implementations of these examples can be found in \citep{hoffman2020acme}, which describes Acme, a package for distributed reinforcement learning that makes heavy use of Launchpad.

\begin{listing}
\begin{python}
# Create an empty program graph.
p = Program('actor-learner')

# Add a node for the Learner.
with p.group('learner'):
  learner = p.add_node(CourierNode(Learner, batch_size=128))

# Add a node for each Actor.
with p.group('actor'):
  for _ in range(num_actors):
    p.add_node(CourierNode(Actor, learner))
\end{python}
\caption{A Launchpad program for a distributed on-policy actor-critic algorithm. This setup separates data collection (interacting with the environment) from parameter updates, and allows us to speed up data collection by adding more actor nodes.}
\label{lst:actor-learner-program}
\end{listing}

\section{Discussion}

Unlike Ray, Launchpad does not itself implement mechanisms providing fault tolerance. Instead, we assume only that the underlying job scheduling system (e.g.\ Kubernetes) has the ability to restart failing jobs. In practice, this requires that stateful nodes---often learning nodes holding model parameters---have the ability to restore themselves. However, we argue that, rather than more ``exact'' recovery mechanisms (such as the lineage recovery used in Ray) this approach is sufficient in practice for many machine learning systems.

For example, in the Actor-Learner reinforcement learning example given earlier it is possible that the actor processes crash. These processes, which are stateless (given the model parameters), do not require checkpointing to recover as they can be restarted with no loss in performance. For learning processes this is not the case, but periodic checkpointing allows us to restart the process with minimal losses in performance. However, by not requiring exact mechanisms for fault recovery we are able to do away with the overhead both in terms of implementation (complexity of introducing fault tolerance semantics in the programming model) and runtime (tracing back and restoring the exact states). This is particularly important when the update steps, as is often the case in machine learning, are frequent and it is only their aggregate behavior that we are interested in.

It is also worth noting that Launchpad discourages the use of cyclic communication, which is reflected in the programming interface: to obtain a handle, the corresponding node must be created first. Although this is discouraged---primarily for reasons of clarity---it is not the case that such models are impossible. Models of this form can be constructed in Launchpad by first constructing an ``empty'' node and configuring it after any necessary nodes consuming its handle have been constructed. This is equivalent in other programming languages to distinguishing between allocation and initialization.

\section{Conclusion}

In this work we have introduced Launchpad, a programming model for distributed machine learning research. Our aim with Launchpad is to provide a simple, expressive framework for specifying distributed systems that is easy to use by machine learning researchers and practitioners. We have also shown, through a number of examples, how Launchpad simplifies the design process of common machine learning algorithms and components. While these examples have been simplified to fit within the constraints of this paper, we believe these approaches to be indicative of the types of programs that can be created using this framework. Ultimately we believe that Launchpad can provide a solid base for simplifying the construction of current machine learning techniques as well as to aid in the designing of more complicated future distributed learning algorithms.

\section{Acknowledgements}

Thanks to Lorenzo Blanco, Dan Horgan, Dan Belov, Nando de Freitas, Bobak Shahriari, Albin Cassirer, and Toby Boyd as well as the developers and contributors of Reverb and TF-Agents for all the helpful discussion around this work.

\bibliographystyle{apalike}
\bibliography{references}

\newpage
\clearpage
\appendix

\section{Additional examples}

\paragraph{Parameter Server}
\label{appendix:examples:parameter_server}

In Listing~\ref{lst:pycacher} we include an example implementation of a \py{Cacher} object which can be used in conjunction with a \py{CourierNode} to construct a caching layer.

\begin{listing}
\begin{python}
class Cacher:
  def __init__(self, client, timeout):
    # Get method names exposed by client
    for name in list_methods(client):
      # client_method makes an RPC call
      client_method = getattr(
          client, name)
      
      # Memoize the return value of
      # client_method for timeout seconds.
      memoized_method = memoize_decorator(
          client_method, timeout=timeout)

      # Use memoized_method instead of
      # client_method.
      setattr(self, name, memoized_method)
\end{python}
\caption{A Python implementation of a caching layer.}
\label{lst:pycacher}
\end{listing}

\paragraph{Map Reduce}
In Listing~\ref{lst:mapper-reducer} we include example implementations of the map and reduce functions introduced in Section~\ref{sec:map-reduce}.

\begin{listing}
\begin{python}
class WordMapper:

  def __init__(self,
               infile_path,
               reducers):
    self._infile_path = infile_path
    self._reducers = reducers
    
  def run(self):
    for reducer in self._reducers:
      reducer.mapper_begin()

    with open(self._infile_path) as f:
      for line in f:
        for word in line.split():
          self._send_word(word)

  def _send_word(self, word):
    n = len(self._reducers)
    idx = hash(word) % n
    self._reducers[idx].reduce(word, 1)
    
  def _done(self):
    

class CountReducer:

  def __init__(self, outfile_path):
    self._active_mappers = 0
    self._counter = {}
    self._lock = threading.Lock():
    self._outfile_path = outfile_path
  
  def reduce(self, key, value):
    with self._lock:
      if item not in self._counter:
        self._counter[key] = value
      else:
        self._counter[key] += value
        
  def mapper_begin(self):
    with self._lock:
      self._active_mappers += 1
      
  def mapper_done(self):
    with self._lock:
      self._active_mappers -= 1
      if self._active_mappers == 0:
        self._done()
  
  def _done(self):
    with open(self._outfile_path, 'a'):
      for key in self._counter:
        count = self._counter[key]
        s = '{} {}'.format(key, count)
        f.write(s)
\end{python}
\caption{An implementations of the map and reduce functions introduced in Section~\ref{sec:map-reduce}.}
\label{lst:mapper-reducer}
\end{listing}

\paragraph{Evolution Strategies}
Included in \lstname~\ref{lst:evolver} are sample implementations of the \py{Evolver} and \py{Evaluator} classes used to implement evolutionary strategies in Section~\ref{sec:evolution}.

\begin{listing}
\begin{python}
class Evolver:
  def __init__(self, evaluators):
    self._evaluators = evaluators
    # Parameters for search distribution
    self._search_dist = init_dist()
         
  def run(self):
    fitnesses = None
    while not should_stop(
        self._search_dist, fitnesses):
      fitnesses = self._gradient_step()

  def _gradient_step(self):
    fitness_futures = []
    for evaluator in self._evaluators:
      # Sample from search distribution.
      params = sample_params(
          self._search_dist)

      # Send sampled parameters evaluator.
      # Returns a future to run the
      # evaluators in parallel.
      f = evaluator.futures.evaluate(
          params)
      fitness_futures.append(f)
    
    # Wait for evaluators to finish and
    # get the result of their evaluations.
    fitnesses = [f.result()
                 for f in fitness_futures]

    # Calculate search gradient.
    grad = calc_grad(
        self._search_dist,
        param_samples,
        fitnesses)

    # Update the search distribution.
    apply_grad(
        self._search_dist, grad)

    return fitnesses
    
class Evaluator:
  def evaluate(self, params):
    return do_evaluation(params)
\end{python}
\caption{Sample implementation of classes implementing evolutionary strategies.}
\label{lst:evolver}
\end{listing}

\paragraph{Actor-Learner}
In Listing~\ref{lst:actor-learner} we include an example implementation of the \py{Actor} and \py{Learner} classes used within the distributed reinforcement learning example given in Section~\ref{sec:examples:rl}.

\begin{listing}
\label{rl-code}
\begin{python}
class Actor:
  def __init__(self, learner):
    self._learner = learner

  def run(self):
    environment = make_environment()
    policy_fn = make_policy_fn()
    params = self._learner.get_params()

    trajectory = []
    timestep = environment.reset()
    while True:
      action = policy_fn(params, timestep)
      trajectory.append((timestep, action))

      if trajectory_done(trajectory):
        learner.put(trajectory)
        params = self._learner.get_params()
	    trajectory = []

      timestep = environment.step(
          action)

class Learner:
  def __init__(self, batch_size):
    self._batch = Queue(maxlen=batch_size)
    self._update_fn = make_update_fn()
    self._params = init_variables()

  def run(self):
    while True:
      wait_until_full(self._batch)
      self._params, loss = update_fn(
          self._params, self._batch)
      print('Loss: {}'.format(loss))

  def put(self, trajectory):
    self._batch.put(trajectory)

  def get_params():
    return self._params
\end{python}
\caption{Example implementations of an Actor and Learner that can be used in a distributed actor-critic reinforcement learning algorithm. The Actor interacts with an environment, sends trajectories of data to a Learner, and periodically receives new policy parameters from it. A typical Launchpad program for this algorithm includes multiple Actor nodes with a single Learner node. }
\label{lst:actor-learner}
\end{listing}

\end{document}